\documentclass[aps,prx,reprint,twocolumn,superscriptaddress,longbibliography,floatfix]{revtex4-2}

\usepackage{times}
\usepackage{graphics}
\usepackage{amssymb}
\usepackage{amsmath}
\usepackage{mathtools}
\usepackage[colorlinks,linkcolor=blue,anchorcolor=blue,citecolor=blue,urlcolor=red]{hyperref}
\usepackage{booktabs}
\usepackage{bm}
\usepackage{tikz}
\usetikzlibrary{quantikz2}
\usetikzlibrary{shapes.geometric}
\usetikzlibrary{decorations.pathreplacing}
\usetikzlibrary{shapes.geometric}

\newcommand{\casql}{Laboratory of Quantum Information, University of Science and Technology of China, Hefei, Anhui, 230026, China}
\newcommand{\casex}{Anhui Province Key Laboratory of Quantum Network, University of Science and Technology of China, Hefei 230026, China}

\newcommand{\aihf}{Institute of Artificial Intelligence, Hefei Comprehensive National Science Center, Hefei, Anhui, 230088, China}
\newcommand{\origin}{Origin Quantum Computing Technology (Hefei) Co., Ltd., Hefei, Anhui, 230026, China}
\newcommand{\ymsc}{Yau Mathematical Sciences Center, Tsinghua University, Beijing 100084, China}
\newcommand{\bimsa}{Yanqi Lake Beijing Institute of Mathematical Sciences and Applications, Beijing 100407, China}

\begin{document}

\title{Variational quantum simulation of many-body dissipative dynamics on a superconducting quantum processor 
}

\author{Huan-Yu Liu}
\affiliation{\aihf}
\affiliation{\casql}

\author{Tai-Ping Sun}
\affiliation{\casql}
\affiliation{\casex}

\author{Zhao-Yun Chen}
\affiliation{\aihf}

\author{Cheng Xue}
\affiliation{\aihf}

\author{Chao Wang}
\affiliation{\origin}

\author{Xi-Ning Zhuang}
\affiliation{\casql}
\affiliation{\casex}
\affiliation{\origin}

\author{Jin-Peng Liu}
\email{liujinpeng@tsinghua.edu.cn}
\affiliation{\ymsc}
\affiliation{\bimsa}

\author{Wei Yi}
\email{wyiz@ustc.edu.cn}
\affiliation{\casql}
\affiliation{\casex}

\author{Yu-Chun Wu}
\email{wuyuchun@ustc.edu.cn}
\affiliation{\casql}
\affiliation{\casex}
\affiliation{\aihf}

\author{Guo-Ping Guo}
\affiliation{\casql}
\affiliation{\casex}
\affiliation{\aihf}
\affiliation{\origin}

\date{\today}
\begin{abstract} 
Open quantum systems host a wide range of intriguing phenomena, yet their simulation on well-controlled quantum devices is challenging, owing to the exponential growth of the Hilbert space and the inherently non-unitary nature of the dynamics. Here we propose and experimentally demonstrate a variational quantum algorithm capable of scalable simulation of non-unitary many-body dissipative dynamics. The algorithm builds on the framework of linear combination of Hamiltonian simulation, which converts non-unitary dynamics into a weighted sum of unitary evolutions. With the further introduction of a simplified quantum circuit for loss-function evaluation, our scheme is suitable for near-term quantum hardware, with the circuit depth independent of the simulation time. We illustrate our scheme by simulating the collective dynamics of a dissipative transverse Ising model, as well as an interacting Hatano-Nelson model, on the superconducting quantum processor Wukong. Our work underlines the capability of noisy intermediate-scale quantum devices in simulating dissipative many-body dynamics and represents a step forward in exploiting their potential for solving outstanding physical problems.
\end{abstract}

\maketitle

\section{Introduction}

Quantum computation has seen rapid progress over the past few decades, driven by the development of high‑fidelity quantum processors based on platforms such as superconducting qubits \cite{Huang2020supercqc, Reed2012qrc} and trapped ions \cite{Bruzewicz2019trappedion}. 
Early demonstrations of quantum advantage \cite{qadvan1,qadvan2} have underscored the potential of quantum devices in tackling classically intractable problems. 
Various quantum algorithms have also been proposed to speed up computations for applications in chemistry \cite{McArdle2020qcc, Cao2019qcqc}, finance \cite{Herman2023qcf, Orús2019qcf,2025liuqci}, machine learning \cite{Biamonte2017qml, Cerezo2022qml}, as well as the simulation of many‑body systems \cite{Georgescu2014quantumsim, Daley2022practical, Altman2021qsim, Ganainynhp}. Among these, quantum simulation is particularly compelling, as it directly exploits the natural compatibility between quantum hardware and the unitary evolution of isolated quantum systems.

However, quantum systems in the real world inevitably couple to the environment, giving rise to decoherence and relaxation processes \cite{oqsreview, ISAR1994opensystems}. The resulting open quantum systems, often many-body in nature, exhibit a wide range of intriguing phenomena, including exotic out-of-equilibrium dynamics \cite{Christopoulos2023ed, Purkayastha2016oed}, steady-state phases and phase transitions \cite{Rakovszky2024steady, Jung1999pt, Nakanishi2022ptpt}, and quantum criticalities with new universality classes \cite{newclasses}.
Furthermore, the intrinsic non-Hermitian nature of the underlying Liouvillian dynamics lends itself to non-Hermitian descriptions \cite{oqsreview}, which provide a physical background, as well as a practical outlet, to the booming field of non-Hermitian physics \cite{nhpreview, nhpmereview}. 
But simulating dissipative dynamics on quantum computers remains a significant challenge, since the inherent non‑unitary time evolution makes direct mapping to quantum gates highly non‑trivial.

Explicitly, since quantum gates are inherently unitary, simulating non-unitary dynamics on quantum computers requires the incorporation of non-unitary operations, which necessitates specially designed quantum algorithms.
For instance, previous studies have shown that a dilation-based approach can be adopted for the task, at the cost of auxiliary qubits \cite{oqsreview, oqsdilate}. The framework involves embedding the non-Hermitian Hamiltonian into an enlarged Hilbert space by coupling to an auxiliary qubit. 
After performing a global unitary evolution, auxiliary qubits are measured, and the desired non-unitary operations on the remainder of the system are realized by post-selecting the measurement results \cite{Wu2019Observation}. Such a framework has been successfully applied to the study of topological phases \cite{Observation2021observation}, out-of-equilibrium quantum dynamics \cite{kivelä2024quantumsim}, solving differential equations \cite{jin2024sch}, and the investigation of the non-Hermitian skin effect (NHSE) on superconducting quantum processors \cite{Shen2025Observation}. Nevertheless, the dilation-based approach is generally complex, with high simulation costs for noisy intermediate-scale quantum (NISQ) devices \cite{Preskill2018quantumcomputingin}.

Alternatively, it has been theoretically proposed that a non-unitary evolution can be recast as a weighted linear combination of unitary operations. This approach, dubbed the linear combination of Hamiltonian simulation (LCHS) \cite{lchs,lchsinf,lchsappqet,lchsrand}, allows the use of linear combinations of unitaries (LCU)~\cite{lcu} to prepare time-evolved states in open systems, and to evaluate states' overlap through quantum interference techniques. Despite the immense potential of the framework, its implementation on near-term quantum hardware remains difficult.
For instance, state preparation via LCU typically demands deep and complex quantum circuits, whereas quantum-interference-based schemes suffer from circuit-depth growth that scales unfavorably with the evolution time. Furthermore, normalization requirements in such schemes impose additional resource overheads. To date, an experimental demonstration of the scheme is still lacking.

In this work, we propose and experimentally demonstrate a variational quantum algorithm (VQA) \cite{vqa} for scalable simulations of dissipative quantum dynamics. Our scheme 
builds on the LCHS framework, but introduces key modifications to improve its feasibility on near-term quantum devices. In our scheme, the parameterized quantum circuits (PQCs) are employed to prepare the time‑evolved states. Time evolution is then converted into an optimization problem, in which the loss function can be efficiently evaluated using a simplified Hadamard test. Crucially, the overall design of our algorithm renders the circuit depth independent of the simulation time, and the simplified Hadamard test substantially reduces the implementation overhead on noisy hardware. Furthermore, the unitary nature of the parametrized ansatz ensures that the output state is automatically normalized without the need of additional post‑processing.
We experimentally demonstrate the quantum simulation algorithm on the superconducting quantum computer Wukong \cite{wukong}, using a dissipative transverse Ising model and an interacting Hatano-Nelson (HN) model as examples. For the latter in particular, we observe the many-body NHSE and its dynamic particle-hole symmetry, both matching theoretical predictions with high fidelity.
Our work marks a significant step in leveraging quantum computing for the study of larger-scale and more realistic physical systems.

\section{Results}\label{sec:results}

\subsection{Algorithm for the variational quantum simulation}

We consider the dissipative dynamics under a time-independent non-Hermitian Hamiltonian $H$. While itself a central topic in non-Hermitian physics \cite{nhp}, the non-unitary state evolution can also be mapped to that of the vectorized density matrix in an enlarged Hilbert space \cite{oqsreview}. Hence, though we focus on the non-Hermitian Hamiltonian $H$ in the following, the simulation scheme is applicable to general open quantum systems.

Our scheme consists of two main steps. First, we express the non-unitary dynamics as a sum of unitary evolutions, following the LCHS approach \cite{lchs}. Specifically, we write the non-Hermitian Hamiltonian as $H = H_0 + iV$, where $H_0 = (H + H^\dagger)/2$ and $V = (H - H^\dagger)/2i$. Under the condition $-V \succeq 0$, the evolution operator $e^{-iHt}$ is recast into the integral
\begin{equation}
e^{-iHt} = e^{-(iH_0 - V)t} = \int_R \frac{1}{\pi(1 + k^2)} e^{-i(H_0 - kV)t} \, dk,
\end{equation}
where the kernel $1/(1 + k^2)$ decays rapidly, allowing for the truncation $k \in [-K, K]$ with controllable error dictated by the cutoff $K$ \cite{lchs}. Discretizing the integral range into steps of $\delta k$, we have
\begin{equation}\label{eq:lchs}
e^{-iHt} \approx \sum_k c_k U_k,
\end{equation}
where $k \in \{-K, -K + \delta k, \dots, K\}$, $c_k = \delta k/\pi(1 + k^2)$, and $U_k = e^{-i(H_0 - kV)t}$. We further set the initial time of the evolution $t_0=0$, and discretize the evolution into steps of $\delta t$, so that the time-evolved state at the $m$th step ($m\in \mathbb{N}$) is given by $|\psi_m\rangle = \frac{1}{C} e^{-iHt} |\psi_0\rangle = \frac 1C (\sum_k c_k U_k)|\psi_0\rangle$, where $t=m\delta t$ and $C$ is a normalization factor.

\begin{figure*}
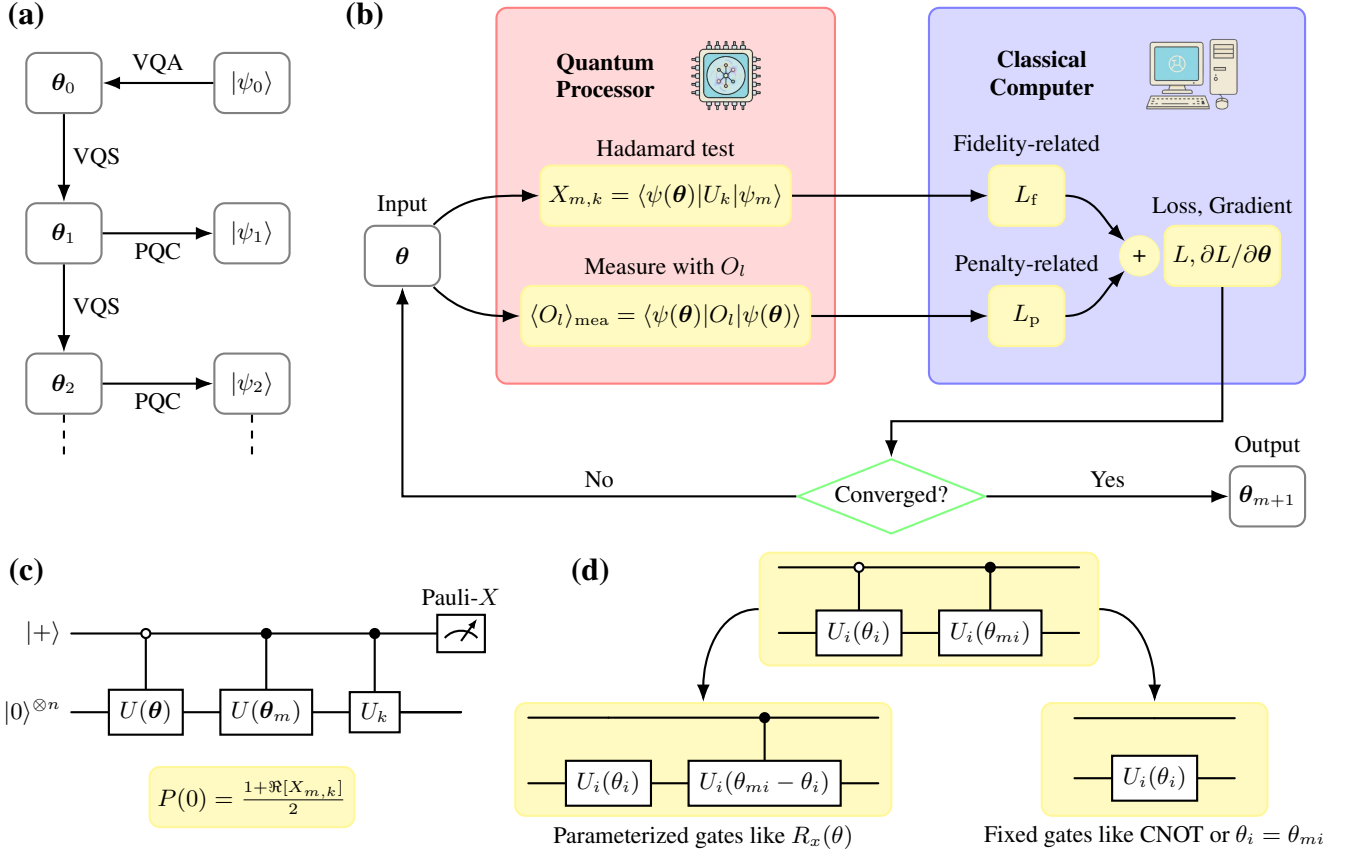

    \centering
    \include{workflow}
    \caption{
    \textbf{Workflow of the variational quantum simulation algorithm.}  
    (a) Schematic overview. The objective is to determine a sequence of parameters $\{\bm{\theta}_m\}$ such that $U(\bm{\theta}_m)\ket{0} = \ket{\psi_m}$. 
    We first use a VQA to obtain $\bm{\theta}_0$ by maximizing the fidelity between $U(\bm{\theta})\ket{0}$ and the initial state $\ket{\psi_0}$. The VQS procedure is then applied to generate the parameters step-by-step.  
    (b) The VQS framework at step $m$. The initial parameters $\bm{\theta}$ are input to a quantum processor. Given the LCHS decomposition $\sum_k c_k U_k$ and a set of observables $\{O_l\}$, we apply the Hadamard test to obtain $X_{m,k}$, and perform direct measurements to evaluate $\langle O_l \rangle_{\operatorname{mea}}$. These quantities are sent to a classical computer to evaluate the loss function and its gradient information. Parameters are optimized in such a hybrid quantum-classical loop until convergence is reached. We then obtain $\bm{\theta}_{m+1}$, and the time-evolved state at the next time step.  
    (c) Quantum circuit for the Hadamard test to evaluate $\Re[X_{m,k}]$. 
    (d) Schematic overview of the simplification procedure. The circuit can be simplified in pairs.
    Further details of the VQS algorithm and the simplification procedure can be found in the Methods section.
    }
    \label{fig:workflow}
\end{figure*}

We then introduce a hybrid variational quantum simulation (VQS) framework to implement the LCHS-based non-unitary evolution. Our VQS algorithm, illustrated in Fig.~\ref{fig:workflow}(a), utilizes a PQC to generate time-evolved states through a sequence of parameters $\{\bm{\theta}_m\}$, such that $U(\bm{\theta}_m)\ket{0} = \ket{\psi_m}$, where $|0\rangle$ denotes the initial computational basis state of all qubits.

For this purpose, we first obtain $\bm{\theta}_0$ by maximizing the fidelity between the ansatz $ | \psi(\bm{\theta})\rangle =  U(\bm{\theta})\ket{0}$ and the initial state $\ket{\psi_0}$. The rest of the parameters are then generated iteratively, as illustrated in Fig.~\ref{fig:workflow}(a)(b). Specifically, given the time-evolved state $\ket{\psi_{m}}$, the parameter $\bm{\theta}_{m+1}$ is obtained by maximizing the fidelity between the ansatz state $\ket{\psi(\bm{\theta})}$ and the target state $\ket{\psi_{m+1}} = e^{-iH\delta t} \ket{\psi_{m}}$. Following Eq. (\ref{eq:lchs}), we define the fidelity-related loss function $L_{\operatorname{f}}(\bm{\theta},m) = 1 - |\sum_k c_k X_{m,k}|^2$, where $    X_{m,k} = \langle \psi(\bm{\theta})|U_k|\psi_m\rangle$ can be obtained through the Hadamard test on the quantum processor, as shown in Fig. \ref{fig:workflow}(c).  To reduce error accumulation, we further introduce a penalty-related loss function $L_{\mathrm{p}}(\bm{\theta}, m)$. For a given set of observables $\{O_{l}\}$, such as those associated with symmetries or laws of conservation, this penalty enforces the measured expectation values $\langle O_{l} \rangle_{\mathrm{mea}} =\langle \psi(\bm{\theta}) | O_{l} | \psi(\bm{\theta}) \rangle$ to match the target values $\langle O_{l} \rangle_{\mathrm{tar}}$. Note that the penalty term, whenever applicable, is particularly helpful in increasing the efficiency of simulating non-Hermitian dynamics, where state norms are generally not conserved. The combined loss function $L(\bm{\theta},m) = L_{\operatorname{f}}(\bm{\theta},m) + L_{\operatorname{p}}(\bm{\theta},m)$ and the associated gradient information are then evaluated on a classical computer, following which the parameters $\bm{\theta}$ are updated. The process above is repeated till convergence, so that the evolved state at the $(m+1)$th step is obtained through $\ket{\psi_{m+1}} = U(\bm{\theta}_{m+1}) \ket{0}$. An essential advantage of our scheme is that the circuit depth remains constant with respect to the simulation time and increases only mildly with the system size. This makes our approach particularly suitable for NISQ devices.

A key element of the VQS scheme is the evaluation of the fidelity-related loss function $L_{\operatorname{f}}  (\bm{\theta},m)$. As shown in Fig. \ref{fig:workflow}(c), the process involves a Hadamard test, where an auxiliary qubit is prepared in the superposition state $|+\rangle = (\ket{0} + \ket{1})/\sqrt{2}$. A sequence of controlled operations is then applied to the system, producing two distinct states conditioned on the auxiliary qubit. Interference between the two states is enabled by measuring the auxiliary qubits in the Pauli-$X$ basis. Here the conventional Hadamard test requires three controlled operations, which generally involve three-qubit gates such as the Toffoli gates. These gates substantially increase the circuit depth and thereby suffer from limited feasibility on near-term quantum processors. As a remedy, we implement an ingeniously designed simplification of the Hadamard test, as shown in Fig. \ref{fig:workflow}(d). While $U(\bm{\theta})$ and $U(\bm{\theta}_m)$ share the same structure, we rearrange the first two controlled operations in Fig. \ref{fig:workflow}(c) in pairs, and each pair can be simplified based on the parameterization. The last $|1\rangle$-controlled-$U_k$ can be processed in a similar manner. Using our simplification procedure, the circuit depth is greatly reduced, with all three-qubit gates eliminated. We note that such a design is general and would facilitate the implementation of algorithms with Hadamard tests on near-term devices.

\subsection{Simulating the dissipative Ising model}

We proceed to implement the VQS scheme above on the superconducting quantum computer Wukong \cite{wukong}. As a first benchmarking example, we consider the collective spin dynamics in a one-dimensional Ising model under a complex transverse magnetic field. The non-Hermitian Hamiltonian is
\begin{equation}
    H = -J \sum_{i=1}^{n-1} Z_i Z_{i+1} - g \sum_{i=1}^n X_i,
\end{equation}
where $n$ is the system size, $J$ denotes the spin--spin coupling strength, and $g = g_{\mathrm{r}} + i g_{\mathrm{i}}$ is the complex magnetic field. In this study, we set $n = 6$, $J = 1$, $g_{\mathrm{r}} = 2$, and impose an open boundary condition. The system is initialized as $\lvert 0\rangle^{\otimes n}$, and we study the time evolution of the average spin polarization $S_z = \frac{1}{n} \sum_{j=1}^n Z_j$ through its expectation value $\langle S_z \rangle$.

\begin{figure}[h]
    \includegraphics[width=1\linewidth]{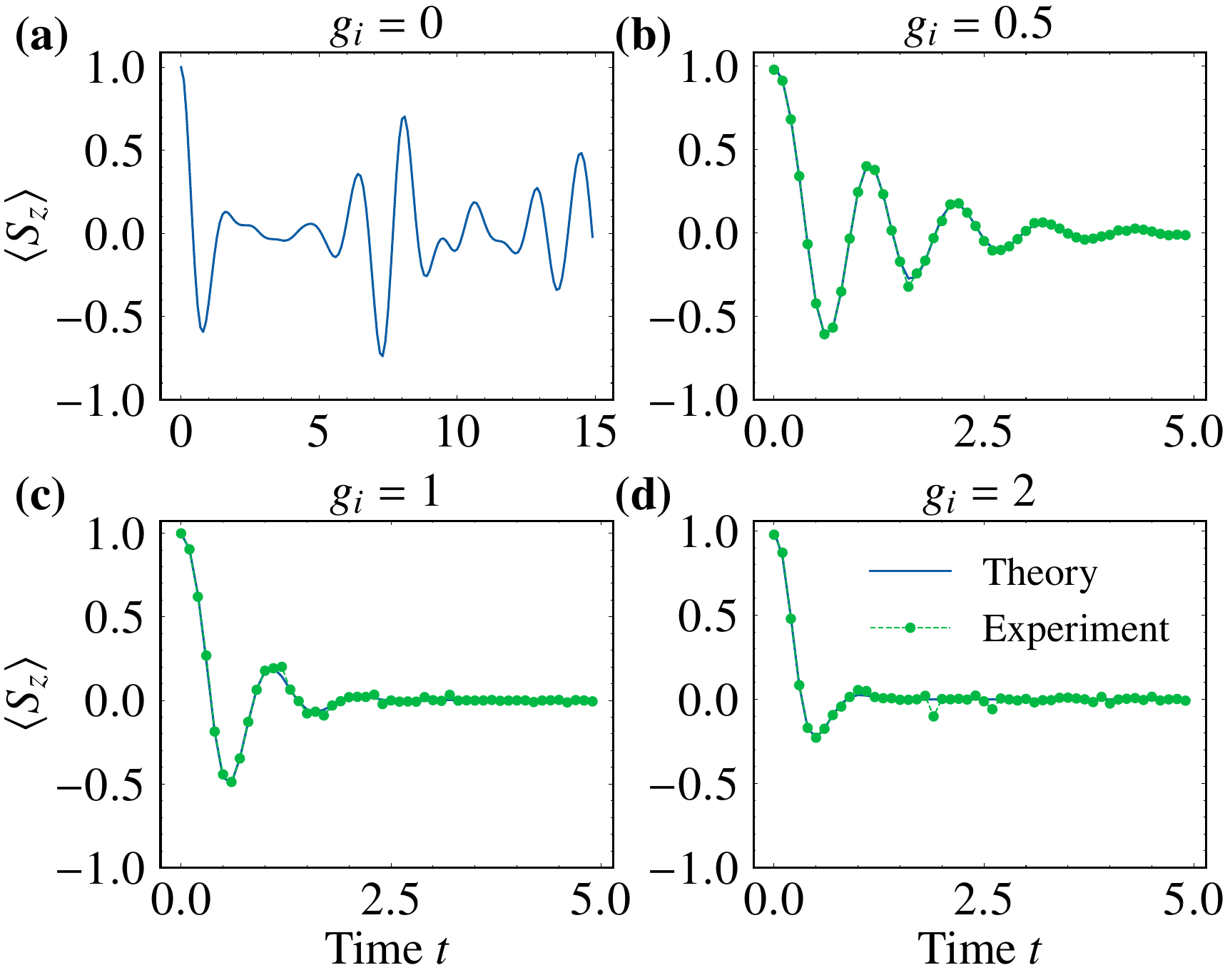}
    \caption{
    \textbf{Simulating results of the dissipative Ising model.} 
    Panels (a)–(d) correspond to $g_{\mathrm{i}} = 0$, $0.5$, $1.0$, and $2.0$, respectively. 
    The solid lines represent results from numerical simulations, and the dots are experimental data.
    For each data point in panels (b)–(d), the parameterized circuit is executed 50 times, with each execution sampled by $2\times 10^{4}$ shots. 
    Among these, the result exhibiting the minimal bias is selected for presentation.
    }
    \label{fig:ising}
\end{figure}

We first examine the evolution of $\langle S_z\rangle$ with increasing $g_{\mathrm{i}}$. For $g_{\mathrm{i}} = 0$, the model reduces to the conventional transverse-field Ising model, wherein $\langle S_z\rangle$ should exhibit persistent non-decaying oscillations over long times [see Fig. \ref{fig:ising}(a)]. Introducing a nonzero imaginary component of the magnetic field ($g_{\mathrm{i}} \neq 0$) would qualitatively change the dynamics, as we expect the magnetization to display damped oscillations.

Experimentally, we focus on the damping dynamics with $g_{\mathrm{i}} \neq 0$. For each given $g_{\mathrm{i}}$, we fix the total evolution time $T = 5$, with the time step $\delta t = 0.1$. We also take $K = 80$ and $\mathrm{d}k = 1$ as the LCHS parameters. The PQC is chosen to be a 5-layer hardware-efficient ansatz, and we only apply the fidelity-related loss function due to the lack of a convenient conserved quantity. Equipped with the optimized parameters from a classical simulator, we then carry out the PQC on the superconducting quantum processor to measure $\langle S_z\rangle$ over time. At each time step, the circuit is run 50 times, each sampled with $2\times 10^4$ measurements. 

Figures~\ref{fig:ising}(b)--(d) compare the experimental results (dots), which are selected with the minimum bias, with the numerical simulations (solid lines). The measured dynamics show excellent agreement with the theoretical predictions.
More importantly, from the mean and standard deviations of the data points (see Supplementary Information), we conclude that the PQC yields reliable simulation results at the current noise level. Consequently, the measurement-based error mitigation built into the quantum processor is sufficient for the simulation task.

\subsection{NHSE and dynamic symmetry of the interacting Hatano--Nelson model}

The NHSE \cite{nhskineff}, where a macroscopic number of eigenstates become exponentially localized at the boundaries of the system, is a hallmark phenomenon in non-Hermitian physics with far-reaching topological and dynamical consequences. To date, the phenomenon has been experimentally demonstrated in both classical \cite{nhsec1,nhsec2,nhsec3} and quantum systems \cite{nhseq1,nhseq2}, but mostly on the single-particle level. In a recent experiment, many-body non-Hermitian Hamiltonians hosting NHSE were implemented on a digital quantum simulator using the dilation-based approach \cite{Shen2025Observation}, but the simulation cost can be challenging due to the complexity of the dilated Hamiltonian.

Here we illustrate the utility of our VQS scheme by simulating the interacting HN model, which promises more scalable simulations in the near term. We focus on the Hamiltonian
\begin{equation}
    H_{\text{p}}(t_R,t_L) = \sum_{j=0}^{n-2} \big( t_R\, c_{j+1}^\dagger c_{j} + t_L\, c_{j}^\dagger c_{j+1} + U\, n_j n_{j+1} \big),
\end{equation}
where the subscript ``p" denotes the particle representation, $c_j^\dagger$ ($c_j$) are fermionic creation (annihilation) operators, $n_j = c_j^\dagger c_j$ is the particle-number operator, $t_R = e^{g}$ and $t_L = e^{-g}$ represent asymmetric right- and left-hopping amplitudes, and $U$ is the nearest-neighbor interaction strength.

The interacting HN model above features a dynamic symmetry, which is related to the particle-hole transformation $c_j \rightarrow (-1)^j c_j^\dagger$. Under such a transformation, the Hamiltonian becomes 
\begin{align}
    &H_{\text{h}}(t_L,t_R) \nonumber\\
    =& \sum_{j=0}^{n-2} \left[ t_L c_{j+1}^\dagger c_{j} + t_R c_{j}^\dagger c_{j+1} + U (1-n_j)(1-n_{j+1}) \right] .
\end{align}
Note that, apart from constants that can be absorbed into the global phase (assuming a fixed filling), the original form of $H_p$ can be recovered from $H_{\operatorname{h}}$, by switching the hopping parameters $t_L\leftrightarrow t_R$. 
We thus expect the particle dynamics under $H_p(t_R,t_L)$ to be the same as the corresponding hole dynamics under $H_{\operatorname{p}}(t_L,t_R)$. 

\begin{figure*}[t]
    \centering
    \includegraphics[width=\linewidth]{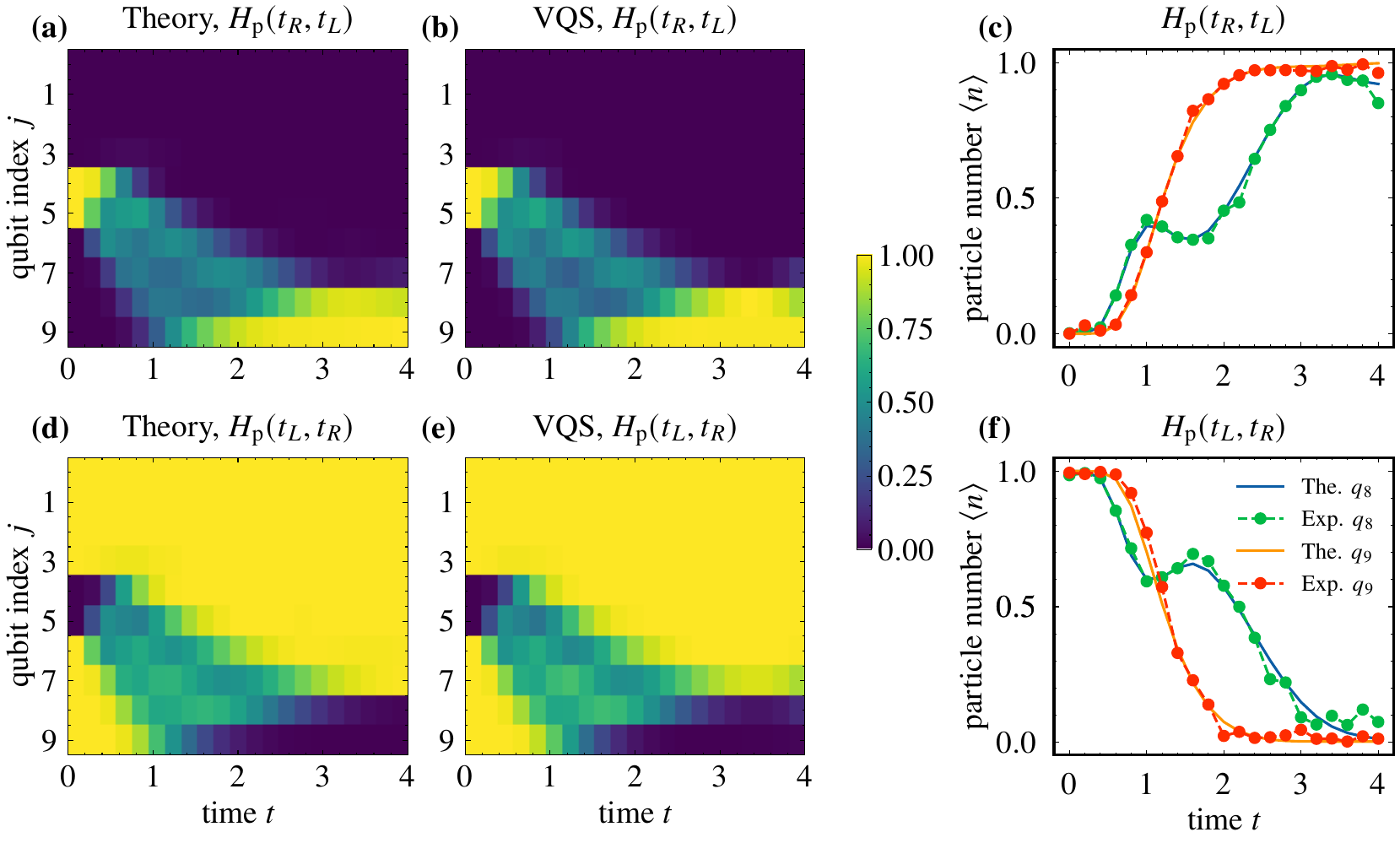}
    \caption{
        \textbf{Simulation results of the interacting HN model.}
        (a) Theoretical predictions of $\langle n_j\rangle$ for $H_{\text{p}}(t_R,t_L)$, showing progressive localization of fermions toward the right edge.  
        (b) Simulation results for $H_{\text{p}}(t_L,t_R)$ using our VQS algorithm, showing excellent agreement with the theoretical results in (a).  
        (c) Experimental evolution of the occupations for the two edge sites $q_8$ and $q_9$ for $H_{\text{p}}(t_R,t_L)$. Each data point is obtained from $2\times 10^4$ shots. The experimental results match the theoretical predictions throughout the time evolution.
        (d)(e)(f) Simulations results for $H_{\text{p}}(t_L,t_R)$.
    }
    \label{fig:hn_results}
\end{figure*}

We perform numerical simulations for $g = 1$, $U = 1$, and a system size of $n = 10$. For evolution under $H_p(t_R,t_L)$, we choose the initial state $|\psi_0\rangle = |0000110000\rangle$ where the central two sites are occupied by fermions. 
Whereas for that under $H_p(t_L,t_R)$, the particle-hole transformed initial state $|\psi_0\rangle = |1111001111\rangle$ is taken. The dynamics are simulated up to a maximum time $T = 4$ with the time step $\delta t = 0.2$, while we record the on-site occupation $n_j(t)$. The theoretically predicted evolutions under $H_p(t_R,t_L)$ and $H_{p}(t_L,t_R)$ are shown in Fig. \ref{fig:hn_results}(a)(d), respectively, where clear signatures of the NHSE are observed. The fermions in Fig. \ref{fig:hn_results}(a) and holes in Fig. \ref{fig:hn_results}(d)
propagate asymmetrically toward the right boundary, and by $t \approx 3$ the density is strongly localized at the edge.

For our variational algorithm, we set $K = 100$, $\delta k = 0.5$, and employ a 5-layer hardware-efficient ansatz as the PQC. Consider the total particle operator $O = \sum_j n_j$. Since $[H,O] = 0$, it follows that $[O,e^{-iHt}] = 0$. For our chosen initial state for $H_p(t_R,t_L)$, $O|\psi_0\rangle = 2|\psi_0\rangle$. By defining $T=e^{-iHt}$, we have
\begin{equation}
    \langle O\rangle_t = \frac{
    \langle \psi_0 | T^\dagger O T|\psi_0\rangle
    } { \langle \psi_0 | T^\dagger T  |\psi_0\rangle} 
    = \frac{
    \langle \psi_0 |T^\dagger T O|\psi_0\rangle
    } { \langle \psi_0 | T^\dagger T |\psi_0\rangle} 
    = 2.
\end{equation}
Hence, the total particle number remains invariant under the normalized non-Hermitian dynamics. We thus choose $O$ for the penalty-related loss function. As in previous sections, parameters are optimized using a classical optimizer. The simulation results using the VQS algorithm are shown in Fig. \ref{fig:hn_results}(b)(e), which match with the theoretical predictions in Fig. \ref{fig:hn_results}(a)(d) with high fidelity.

For our experiment, we implement the optimized PQCs and measure the occupations of the two rightmost qubits, $q_8$ and $q_9$. For each time step, the circuit is executed with $2\times 10^4$ shots. As shown in Fig. \ref{fig:hn_results}(c)(f), experimental data from the quantum processor agree well with theoretical predictions. Note that occupations on other sites can be experimentally obtained in a similar fashion.

\section{Discussion}\label{conclusion}

We theoretically propose and experimentally demonstrate a general variational quantum simulation algorithm for many-body dynamics in open quantum systems. Our scheme builds on the LCHS framework, allowing for the simulation of nonunitary dynamics through a series of unitary evolutions. The experimental implementation is further facilitated by a novel design of the Hadamard test, which significantly reduces the complexity of 
the quantum circuit employed to evaluate the cross terms of the loss function. Our work illustrates that VQAs can be adapted to capture the rich dynamics of open quantum systems on current hardware, paving the way for experimental studies of phenomena that are beyond the reach of classical computation.

While our simplification procedure improves the feasibility of our scheme on noisy quantum devices, it can still be resource-intensive in practice. As discussed in Ref. \cite{Chakraborty2024implementingany}, in certain cases, the loss function can be approximated by retaining only the diagonal terms, which can be measured more efficiently without auxiliary qubits. This approach may be beneficial when circuit depth or qubit count is severely constrained. Another important consideration is simulation accuracy. The central design principle here is to balance hardware implementation with the training of the VQS. Once this trade-off is under control, further improvements to accuracy can be introduced. For example, Ref. \cite{lchsdec} proposed modified kernel functions beyond $1/(1+k^2)$, which should improve truncation and discretization accuracy. With further hardware advances, one could also optimize the time step $\delta t$ to refine precision. Further, we expect that a judicious selection of the observables in the penalty-related loss function can mitigate the fidelity decay in experiments.

\section{Method}\label{sec:method}

\subsection{Implementation and limitations of the LCHS method}\label{subsec:lchs-methods}

For quantum dynamics, we always track expectation values of an observable $O$. Substituting the expression in Eq.~\eqref{eq:lchs} gives
\begin{equation}\label{eq:expectc}
\langle \psi_m | O | \psi_m \rangle = \frac{1}{C^2} \sum_{k,k'} c_k c_{k'} \langle \psi_0 | U_k^\dagger O U_{k'} | \psi_0 \rangle.
\end{equation}
This admits two implementations. First, the state $|\psi_m\rangle$ can be prepared using LCU-based techniques~\cite{lcu}, followed by direct measurement of $O$. Second, the cross terms $\langle \psi_0 | U_k^\dagger O U_{k'} | \psi_0 \rangle$ (and normalization) can be estimated pairwise via the Hadamard test, with classical post-processing. 

The LCHS approximation in Eq.~\eqref{eq:lchs} requires $-V \succeq 0$. This can always be enforced by shifting $V \to V - xI$ with $x$ exceeding the largest eigenvalue of $V$~\cite{lchs}. For an $n$-qubit operator $V = \sum_i v_i P_i$ expressed in the Pauli basis ($P_i \in \{X, Y, Z, I\}^{\otimes n}$, $v_i \in \mathbb{R}$), the eigenvalues of each $P_i$ are $\pm 1$. Thus, for any state $|\phi\rangle$,
\begin{equation}
\langle \phi | V | \phi \rangle = \sum_i v_i \langle \phi | P_i | \phi \rangle \leq \sum_i |v_i|,
\end{equation}
and $x = \sum_i |v_i|$ suffices as a tight bound.

For near-term quantum devices, implementing LCHS poses challenges: LCU requires ancillary qubits and deep circuits for selectives. The depth for each $U_k$ grows linearly with $t$, while Hadamard tests add further overhead from controlled operations. These limitations motivate our hybrid optimizations, detailed below.

\subsection{Details on the VQS Algorithm}

VQA is a hybrid quantum–classical approach designed to solve optimization problems, in which a PQC prepares the ansatz, quantum measurements evaluate loss functions, and a classical optimizer updates the circuit parameters. Although VQAs face certain challenges, such as trainability issues~\cite{bp, bpreview} and reachability concerns~\cite{reach}, the use of relatively shallow PQCs makes them a promising candidate for execution on near-term quantum hardware. 

VQAs have been extensively utilized for quantum simulation. Using the dilation method~\cite{Wu2019Observation}, a VQA was proposed to simulate small-scale non-Hermitian systems, where a PQC approximates the dynamics generated by the dilated Hamiltonian~\cite{Liu2023practical}. A VQS framework \cite{vqs} was developed for general dynamical processes, with specific variational principles mapping the evolution of quantum states to equations of motion for the PQC parameters. Within the VQA framework, dynamical phase transitions in the quantum Ising model have been explored on superconducting quantum processors~\cite{Dborin2022Simulating}.

Here, we provide details about our VQS algorithm. At iteration step $m$, with $\ket{\psi_m} = U(\bm{\theta}_m)\ket{0}$, $\bm{\theta}_{m+1}$ is determined by maximizing the fidelity between the ansatz $\ket{\psi(\bm{\theta})}$ and the target state $\ket{\psi_{m+1}} = e^{-iH\delta t} \ket{\psi_m}$. The fidelity-related loss function, derived from the LCHS-based representation, is defined as
\begin{equation}\label{eq:loss}
\begin{aligned}
    L_{\mathrm{f}}(\bm{\theta},m) 
    &= 1 - \left|\langle \psi(\bm{\theta})| e^{- iH\delta t}|\psi_m\rangle\right|^2 \\
    &= 1 - \left| \sum_k c_k \langle \psi(\bm{\theta})|U_k|\psi_m\rangle \right|^2 .
\end{aligned}
\end{equation}
A multiplicative normalization constant $C$ may be omitted, as it does not change the stationary point.

Approximation errors may accumulate over successive steps. This effect is particularly pronounced for non-Hermitian dynamics, where the evolution is non-unitary and the state norm is not preserved. To suppress such errors, we introduce a penalty term that enforces physical constraints:
\begin{equation}\label{eq:losspea}
    L_{\mathrm{p}}(\bm{\theta},m) = \lambda \sum_l \left( \langle O_l\rangle_{\operatorname{mea}} - \langle O_l\rangle_{\operatorname{tar}} \right)^2 ,
\end{equation}
where $\lambda > 0$ is the penalty coefficient, $\langle O_l\rangle_{\operatorname{mea}}  = \langle \psi(\bm{\theta})|O_l|\psi(\bm{\theta})\rangle$ is the measured expectation value, and $\langle O_l\rangle_{\operatorname{tar}} $ is the corresponding target value.  Then the total loss function is given by
\begin{equation}
    L(\bm{\theta},m) = L_{\mathrm{f}}(\bm{\theta},m) + L_{\mathrm{p}}(\bm{\theta},m) .
\end{equation}
The optimal parameters are obtained as $   \bm{\theta}_{m+1} = \arg\min_{\bm{\theta}} L(\bm{\theta},m)$, which define the next state $\ket{\psi_{m+1}} = U(\bm{\theta}_{m+1})\ket{0}$.

Besides designing suitable loss functions, improving performance in the optimization process is also crucial for achieving high accuracy. Compared with randomly initialized parameters, our previous work \cite{Liu2023mitigating} suggests that transfer-learning-inspired parameter initialization can be beneficial: when optimizing $L(\bm{\theta},m)$, the initial parameters can be set as $\bm{\theta} = \bm{\theta}_m$, which improves both trainability and overall optimization efficiency.

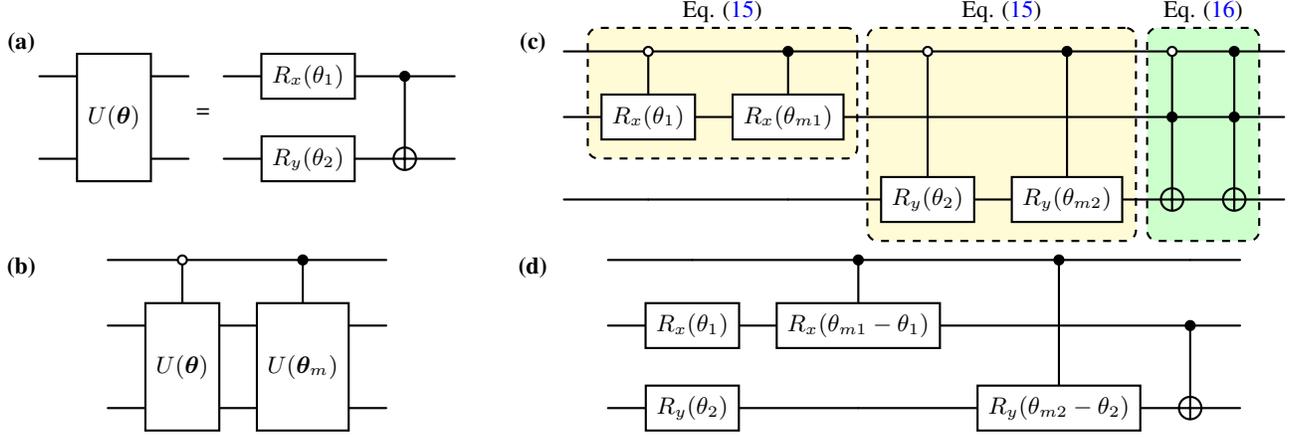
\begin{figure*}[ht]
    \centering
        \begin{tikzpicture}
    \node[scale=1] at (0,0) {
    \begin{tikzcd}
        & \gate[wires=2]{U(\bm{\theta})}& \\
        & &
    \end{tikzcd}=
    \begin{tikzcd}
        & \gate{R_x(\theta_1)} & \ctrl{1} & \\
        & \gate{R_y(\theta_2)} & \targ{} & 
    \end{tikzcd}
    };

    \node[scale=1] at (0,-3){
        \begin{tikzcd}
            & \octrl{1}                         & \ctrl{1}                         & \\
            & \gate[wires=2]{U(\bm{\theta})}  & \gate[wires=2]{U(\bm{\theta}_m)}   & \\
            &                                   &                                  & 
        \end{tikzcd}        
    };

    \node[scale=1] at (9,0){
        \begin{tikzcd}
            & \octrl{1}\gategroup[2,steps=2,style={dashed,
            rounded corners,fill=yellow!20, inner xsep=2pt},
            background]{ Eq. \eqref{eq:simsingle}}                & \ctrl{1}                 & \octrl{2}\gategroup[3,steps=2,style={dashed,
            rounded corners,fill=yellow!20, inner xsep=2pt},
            background]{ Eq. \eqref{eq:simsingle}}                 & \ctrl{2}                 & \octrl{1}\gategroup[3,steps=2,style={dashed,
            rounded corners,fill=green!20, inner xsep=2pt},
            background]{ Eq. \eqref{eq:simdouble}}  & \ctrl{1} & \\
            & \gate{R_x(\theta_{1})}  & \gate{R_x(\theta_{m1})}     &                          &                          & \ctrl{1}  & \ctrl{1} & \\
            &                          &                          & \gate{R_y(\theta_{2})}  & \gate{R_y(\theta_{m2})}     & \targ{}   & \targ{}  &
        \end{tikzcd}   
    };

    \node[scale=1] at (9,-3){
        \begin{tikzcd}
            &              & \ctrl{1}                 &                  \ctrl{2}            & &\\
            & \gate{R_x(\theta_{1})}  & \gate{R_x(\theta_{m1}-\theta_{1})}     &                                                  & \ctrl{1}  & \\
            &   \gate{R_y(\theta_{2})}                        &                          & \gate{R_y(\theta_{m2}-\theta_{2})}     & \targ{}   &
        \end{tikzcd} 
    };

    \node at (-3,1) {\textbf{(a)}};
    \node at (3.8,1) {\textbf{(c)}};
    \node at (-3,-2) {\textbf{(b)}};
    \node at (3.8,-2) {\textbf{(d)}};

    \end{tikzpicture}
    \caption{
    \textbf{
    Circuit simplification for the Hadamard test with a two-qubit PQC as one example.
    }
    (a) The example of the two-qubit PQC with two single-qubit gates and a two-qubit gate.
    (b) Circuit for $C_1U(\bm{\theta})\cdot C_0U(\bm{\theta}_m)$.
    (c) Detailed quantum circuit for (b) after we rearrange the quantum gates according to Eq. \eqref{eq:arrange}, which has 4 two-qubit gates and 2 three-qubit gates. Quantum gates in yellow and green dashed boxes can be simplified similarly to Eqs. \eqref{eq:simsingle} and \eqref{eq:simdouble}, respectively.
    (d) The simplified quantum circuit. Compared to (c), the circuit depth is reduced, and there are no three-qubit gates.
    }
    \label{fig:simplify}
\end{figure*}

\subsection{Details on the Hadamard test and its simplification}

The fidelity-related loss function can be evaluated by the Hadamard test shown in Fig. \ref{fig:workflow}(c). In the procedure, an auxiliary qubit is prepared in the superposition state $(\ket{0} + \ket{1})/\sqrt{2}$. Controlled operations are then applied to the system, producing two distinct states conditioned on the auxiliary qubit value. In the circuit shown in Fig.~\ref{fig:workflow}(c), these states are $\ket{0}\otimes U(\bm{\theta})\ket{0}$ and $\ket{1}\otimes  U_k U(\bm{\theta}_m)\ket{0}$. Measuring the auxiliary qubit in the Pauli-$X$ basis is realized by a Hadamard gate, whicht enables interference between the two branches, yielding a joint state of the form $\ket{0} \left( a\,U(\bm{\theta})\ket{0} + b\,U_k U(\bm{\theta}_m)\ket{0} \right)$, with $a$ and $b$ amplitudes. Then we have $P(0) = \left\| a U(\bm{\theta})\ket{0} + b U_k U(\bm{\theta}_m)\ket{0} \right\|^2$, from which the real part of the overlap $\Re[X_{m,k}]=2P(0)-1$ can be extracted. To evaluate the imaginary part, we can apply a phase gate after the first Hadamard gate. 

Here, we show that the Hadamard test can be simplified and the circuit depth can be greatly reduced. An overview of a two-qubit example of our simplification procedure is shown in Fig. \ref{fig:simplify}.

Denote the gate $ C_x U ,  x \in \{0,1\} $ as the controlled operation that applies the unitary $ U $ to the target system when the control qubit is in the state $ \ket{x} $. For example, in this notation, the CNOT gate is written as $ C_1 X $. Our simplification procedure begins with the following two identities:
\begin{align}
   & [C_0 U,\, C_1 V] = 0, \quad \forall\, U, V,  \label{simpcom}\\
   & C_0 U \cdot C_1 U = U. \label{simp01u}
\end{align}

We first focus on $C_1U(\bm{\theta}_m)\cdot C_0U(\bm{\theta})$. Let the PQC be expressed as $ U(\bm{\theta}) = \prod_i U_i(\theta_i)$, where each $ U_i $ in the ansatz is either a parameterized single-qubit gate (e.g., $ R_x $ or $ R_z $) or an unparameterized two-qubit gate (e.g., $\mathrm{CNOT}$ or $\mathrm{CZ}$). An example of a two-qubit PQC and the corresponding product are shown in Fig.~\ref{fig:simplify}(a) and (b). With Eq. (\ref{simpcom}), the product can be rearranged as (see Fig.~\ref{fig:simplify}(c)):
\begin{equation}\label{eq:arrange}
\begin{aligned}
    C_1 U(\bm{\theta}_m) \cdot C_0 U(\bm{\theta}) 
    &= \prod_i C_1 U_i(\theta_{mi}) \cdot  \prod_i C_0 U_i(\theta_{i}) \\
    &= \prod_i \left[ C_1 U_i(\theta_{mi}) \cdot C_0 U_i(\theta_{i}) \right].
\end{aligned}
\end{equation}
Then for each $ i $, if $ U_i $ is parameterized like $ R_z(\theta) $, we have:
\begin{equation}\label{eq:simsingle}
\begin{aligned}
     &C_1 R_z(\theta_{mi}) \cdot C_0 R_z(\theta_{i}) \\
     =&  C_1 R_z(\theta_{mi} - \theta_{i}) \cdot C_1 R_z(\theta_{i}) \cdot C_0 R_z(\theta_{i}) \\
     =& C_1 R_z(\theta_{mi} - \theta_{i}) \cdot R_z(\theta_{i}),
\end{aligned}
\end{equation}
which shows that two controlled gates can be reduced to one controlled gate and one single-qubit gate. Othervise, if $ U_i $ is a two-qubit gate such as $ \mathrm{CZ} $, the reduction is even simpler:
\begin{equation}\label{eq:simdouble}
    C_0 \mathrm{CZ} \cdot C_1 \mathrm{CZ} = \mathrm{CZ},
\end{equation}
meaning that two controlled-$\mathrm{CZ}$ gates can be replaced directly with a single $ \mathrm{CZ} $ gate.

By applying the reductions in Eqs.~\eqref{eq:simsingle} and \eqref{eq:simdouble}, illustrated by the yellow and green dashed boxes in Fig.~\ref{fig:simplify}(c), the full quantum circuit can be simplified, as shown in Fig.~\ref{fig:simplify}(d). Compared to Fig.~\ref{fig:simplify}(c), the circuit depth is reduced, and non-three-qubit gates are needed.

We now turn to the implementation of the gate $C_1U_k$ with
$U_k = e^{-i \delta t  (H_0 - kV)}$.
When the Hamiltonian is expressed as a linear combination of Pauli strings, the corresponding quantum circuit typically consists of parameterized $R_z$ gates together with non-parameterized gates such as $H$, $R_x(\pm \pi/2)$, and CNOT \cite{li2022unified}. 

Suppose that in the Hadamard test, prior to applying this gate, the system is in the (generally unnormalized) state $|0\rangle |\phi_0\rangle + |1\rangle |\phi_1\rangle$. As discussed previously, applying $C_1U_k$ followed by a Hadamard gate on the control qubit produces interference between the states $|\phi_0\rangle$ and $U_k|\phi_1\rangle$, thereby enabling the evaluation of the overlap $\langle \phi_0| U_k | \phi_1 \rangle$.

Interestingly, the same overlap can be obtained if one first applies the controlled operation $C_1U_k$ and then $U_k$ to the target register. In this case, the overlap becomes
$\langle \phi_0| U_k^\dagger \, U_k \, U_k | \phi_1 \rangle = \langle \phi_0| U_k | \phi_1 \rangle$,
which shows the equivalence. Therefore, we can rewrite the operation as
$C_1U_k \ \to\  U_k \cdot C_1U_k$.
Following the inverse process of Eq. (\ref{simp01u}), we have
\begin{equation}
    U_k \cdot C_1U_k = C_0U_k \cdot C_1U_k \cdot C_1U_k.
\end{equation}
Then, using the first-order Trotter approximation, we obtain:
\begin{equation}
  [  C_1 e^{-i\delta t (H_0-kV)} ]^2
    = C_1 e^{-2i\delta t (H_0-kV)} + \mathcal{O}(\delta t^2).
\end{equation}
Consequently, up to an approximation error of order $\mathcal{O}(\delta t^2)$, the resulting gate
$
C_1 e^{-i\delta t (H_0 - kV)} \, C_0 e^{-2i\delta t (H_0 - kV)}
$
is obtained. This can be further simplified using the same procedure illustrated in Fig.~\ref{fig:simplify}.

After simplification, the circuit depth is significantly reduced, and all three-qubit quantum gates are eliminated. Although implementing the method remains challenging on current quantum hardware due to limited qubit connectivity, its scalability is nevertheless promising. Moreover, this simplification method can also be applied to related algorithms, including the original LCHS method and others with multiple controlled quantum gates.

\subsection{Hardware parameters and simulation details}

The experiments in this work were performed on the superconducting quantum processor Wukong \cite{wukong}. The corresponding device parameters are summarized in Table~\ref{wukongpara}. In our simulations, we used at most ten qubits in a chain, selected according to their benchmarked fidelities and to reduce the impact of crosstalk. 

\begin{table}[ht]
\caption{Hardware parameters of the superconducting quantum processor \emph{Wukong}.}
\centering
\begin{tabular}{ll}
\toprule
\textbf{Performance metric} & \textbf{Value} \\
\midrule
Number of qubits & 102 \\
Average single-qubit gate fidelity & 99.74\% \\
Average two-qubit gate fidelity & 96.57\% \\
Average $T_1$ & $33.44\,\mu\mathrm{s}$ \\
Average $T_2$ & $6.93\,\mu\mathrm{s}$ \\
\bottomrule
\end{tabular}
\label{wukongpara}
\end{table}

For the PQC, we employ a five-layer hardware-efficient ansatz.  
Each layer consists of:  
(i) a single-qubit rotation block applying $R_x R_z R_x$ with three independent parameters on every qubit, and  
(ii) a two-qubit entangling block applying $CZ$ gates between nearest neighbors.  
Since all $CZ$ gates commute, they can be scheduled in two sequential sublayers.

Parameter optimization is performed using the BFGS algorithm. To maintain the target simulation accuracy, training is terminated only when the loss function drops below $10^{-2}$.

After training the PQC parameters on a classical simulator, we execute the circuits on the quantum processor. Each circuit is run multiple times to obtain reliable statistics.  
In the main text, we show results corresponding to the minimal bias.

\section*{Acknowledgements}
This work has been supported by the National Key Research and Development Program of China (Grant No. 2023YFB4502500).
J.-P. L. acknowledges support from Innovation Program for Quantum Science and Technology (Grant No.2024ZD0300502), start-up funding from Tsinghua University and Beijing Institute of Mathematical Sciences and Applications. W. Y. is supported by the National Natural Science Foundation of China (Grant No. 12374479).

\bibliography{ref.bib}

\clearpage
\setcounter{table}{0}
\renewcommand{\thetable}{S\arabic{table}}%
\setcounter{figure}{0}
\renewcommand{\thefigure}{S\arabic{figure}}%
\setcounter{section}{0}
\setcounter{equation}{0}
\renewcommand{\theequation}{S\arabic{equation}}%

\onecolumngrid

\begin{center}
    {\large \bf Supplementary Information 
    }
\end{center}

\section{Benchmarking of the LCHS Method and Demonstration of Training}

In this section, we present both the performance benchmarking of the LCHS method and the demonstration of our algorithm’s training procedure. We consider the single-qubit non-Hermitian Su–Schrieffer–Heeger model \cite{Observation2021observation,ssh2}, whose Hamiltonian is
\begin{equation}
    H = \gamma \left[ h_x X + \left(h_z + \frac{i}{2}\right) Z \right],
\end{equation}
where $h_x = v + r \cos k$ and $h_z = r \sin k$ are momentum-dependent coefficients, $v$ and $r$ are hopping amplitudes, $k$ is the Bloch wave vector, and $\gamma$ scales the overall energy scale. In this study we set $v = 0.3$, $r = 1$, $\gamma = 3.5$, and $k = 0.3\pi$. The initial state is $|0\rangle$, and we simulate the dynamics over a total evolution time $t = 1.5$ with a time step $\delta t = 0.05$. We monitor the Loschmidt echo \cite{lecho1,lecho2},
$
\langle P_0 \rangle = |\langle 0 | \psi(t) \rangle |^2,
$
throughout the dynamics.

\begin{figure}[h]
    \centering
    \includegraphics[width=0.5\linewidth]{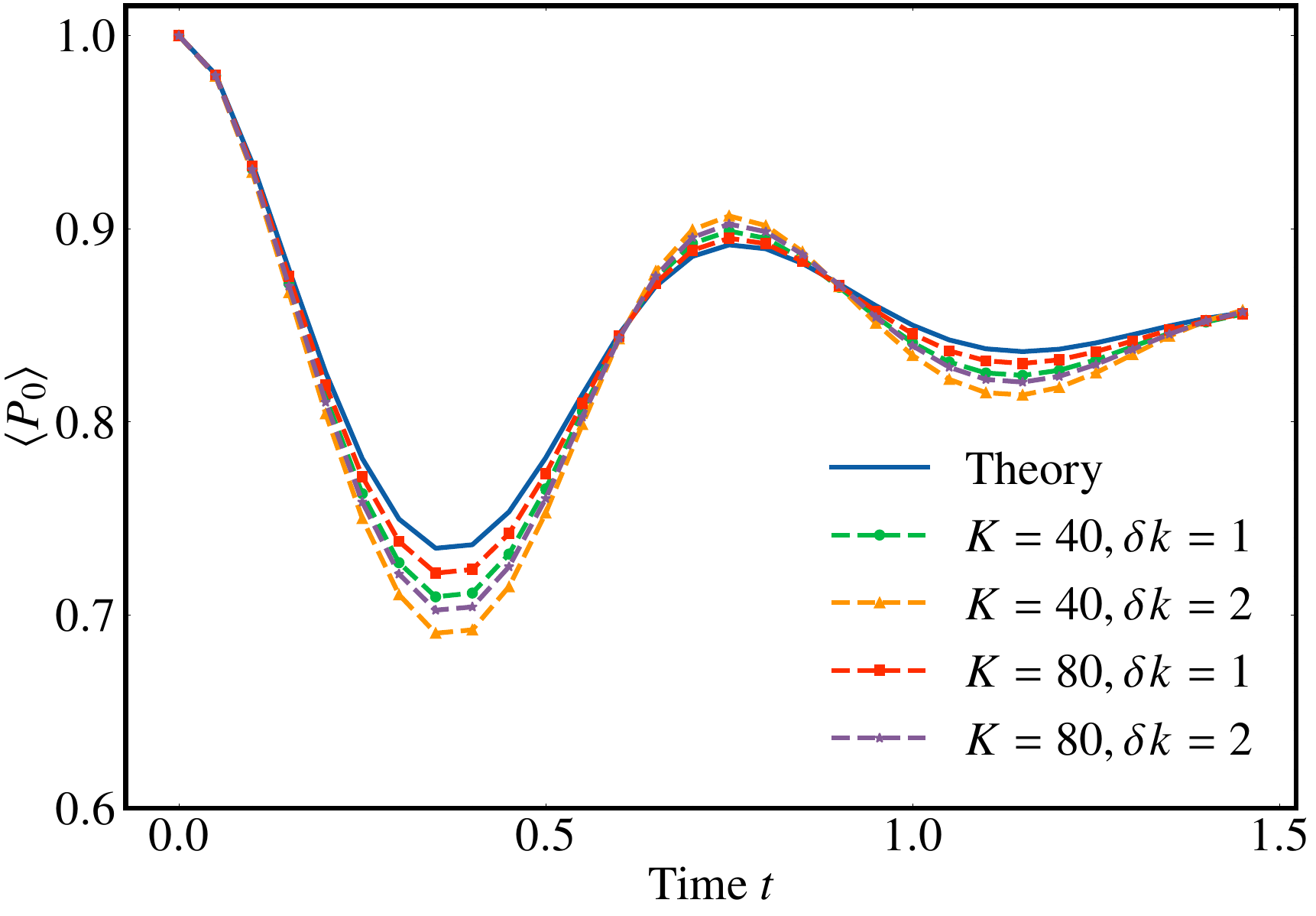}
    \caption{
    \textbf{Benchmarking of the LCHS method for different choices of $K$ and $\delta k$}. 
    Higher accuracy is achieved with larger $K$ and smaller $\delta k$. Notably, $\{K=40, \delta k=1\}$ yields better performance than $\{K=80, \delta k=2\}$ with equal resource cost.
    }
    \label{fig:ssh_bench_lchs}
\end{figure}

We first benchmark the LCHS method using a uniform quadrature to discretize the integral, for various $\{K, \delta k\}$ combinations. The results are shown in Fig.~\ref{fig:ssh_bench_lchs}. As expected, the sampling accuracy improves with increasing $K$ and decreasing $\delta k$. Moreover, the parameter set $\{K=40, \delta k = 1\}$ outperforms $\{K=80, \delta k = 2\}$, despite requiring the same sampling resources. This indicates that the large-$k$ region can be truncated without loss of accuracy, provided the remaining domain is sampled with sufficient resolution.

We then evaluate the variational quantum simulation algorithm using $K = 40$ and $\delta k = 1$. Since the model is single-qubit, a PQC consisting of a single $U_3(\theta, \phi, \lambda)$ gate is already universal. To minimize circuit depth, we implement the entire Hadamard-test operation as a single block and then decompose it into elementary quantum gates. The PQC is initialized as $U_3(0,0,0)$, i.e., the identity, which prepares the $|0\rangle$ state.  

\begin{figure}[h]
    \centering
    \includegraphics[width=0.5\linewidth]{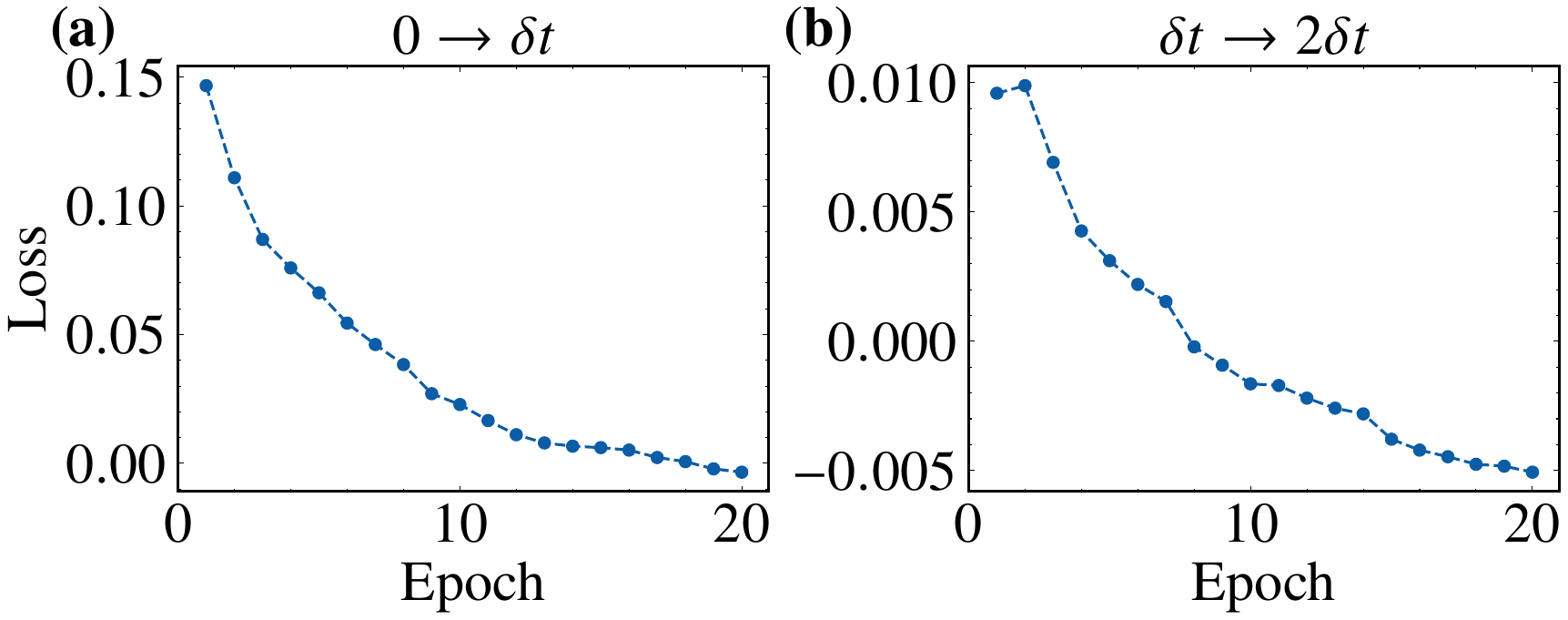}
    \caption{
    \textbf{Optimization of the PQC on quantum hardware during the first two time steps of the simulation.}
    Negative cost values arise from noise in the device.
    }
    \label{fig:sshopt}
\end{figure}

Following this setup, we train the PQC parameters directly on quantum hardware. Figure~\ref{fig:sshopt} shows the optimization trajectory for the first two time steps ($0$–$2\delta t$). The results confirm that optimization proceeds successfully on the device. We note that the cost function occasionally falls below zero, an artifact attributable to hardware noise.

\section{Noise in the quantum hardware}

One important factor affecting the performance of current quantum computers is quantum noise. To mitigate noise on the quantum hardware, we apply a measurement-based error mitigation protocol.  
In this procedure, several calibration circuits are first executed to estimate the probabilities of measurement bit-flips for both logical ``0'' and ``1''.  
Following data acquisition, a classical post-processing correction is applied to obtain the mitigated results.  

\begin{figure}[h]
    \centering
    \includegraphics[width=0.8\linewidth]{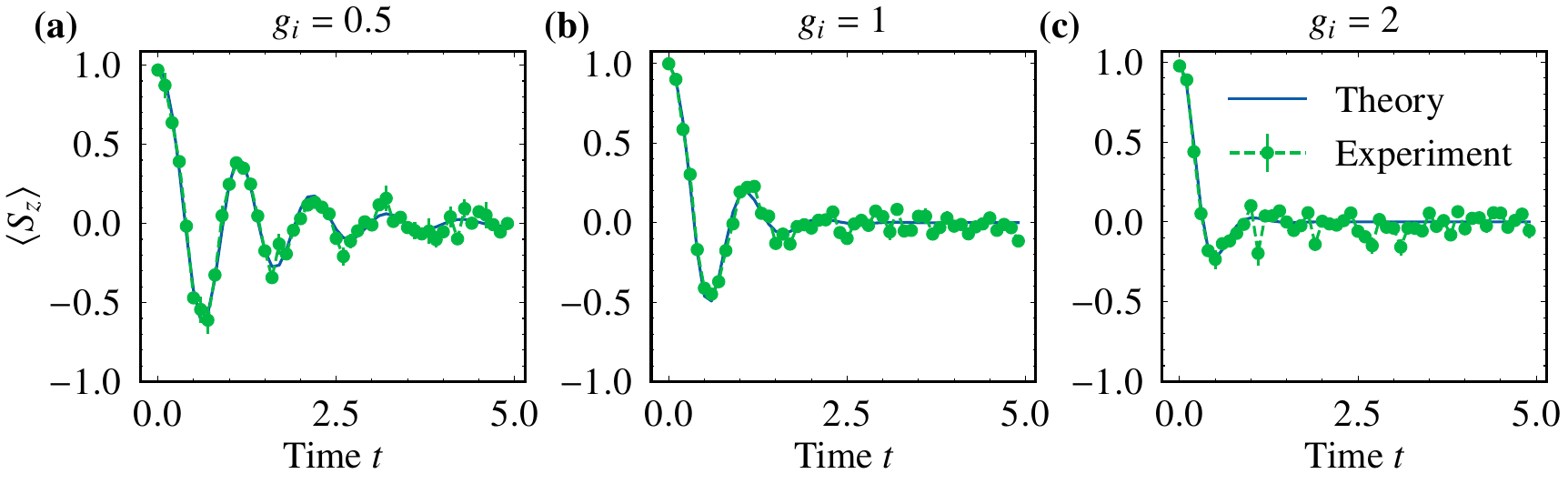}
    \caption{
    \textbf{Extra simulation results for the dissipative Ising model.}
    (a)–(c) correspond to $g_{\mathrm{i}}=0.5$, 1, and 2, respectively. For each point, the circuit is executed 50 times, and the mean value with standard error defines the error bar.
    }
    \label{fig:isingerrorbar}
\end{figure}

In the study of decay dynamics in the Ising model, the PQC is executed 50 times for each parameter point, each of which is measured with $2\times 10^4$ shots. In the main text, we show the experiment results corresponding to the run with minimum bias. Here, we compute the mean value and the associated standard error, and plot the resulting error bars in Fig.~\ref{fig:isingerrorbar}. Panels~(a)–(c) correspond to $g_{\mathrm{i}} = 0.5$, 1, and 2, respectively.  

The results demonstrate that, for the present problem size and noise level, statistical fluctuations are small compared to the overall signal, indicating strong sampling robustness. Consequently, no additional error-mitigation strategies were applied in this study. Future work may explore more advanced error mitigation techniques \cite{qemreview} in the context of larger systems or more complex problem settings.

\end{document}